\documentclass[journal]{IEEEtran}

\usepackage{amsmath,amssymb,amsfonts}
\usepackage{graphicx}
\usepackage{textcomp}
\usepackage{epsfig}
\usepackage{multirow}
\usepackage{float}
\usepackage{algpseudocode, algorithm, tabularx}
\usepackage{algorithmicx}
\usepackage[caption=false]{subfig}
\usepackage{varwidth}
\usepackage{url}
\usepackage{amssymb}
\usepackage{amsthm}
\usepackage{mathtools}
\usepackage{hyperref}
\usepackage{xcolor,cite,etoolbox}

\DeclareMathOperator*{\find}{find}

\newtheorem{lemma}{Lemma}

\newcommand{\etal}{\emph{et al.}}
\newcommand{\cm}{\text{cm}}
\newcommand{\cp}{\text{cp}}

\makeatletter 
\pretocmd\@bibitem{\color{black}\csname keycolor#1\endcsname}{}{\fail}
\newcommand\citecolor[1]{\@namedef{keycolor#1}{\color{blue}}}
\makeatother

\begin{document}

\title{UAV Communications \\for Sustainable Federated Learning}


\author{Quoc-Viet Pham, Ming Zeng, Rukhsana Ruby, Thien Huynh-The, and~Won-Joo Hwang
    \thanks{Q.-V. Pham is with the Korean Southeast Center for the 4th Industrial Revolution Leader Education, Pusan National University, 
    Korea. 
    M. Zeng is with the Department of Electrical Engineering and Computer Engineering, Laval University, 
    Canada.
    R. Ruby is with the College of Computer Science and Software Engineering, Shenzhen University, China.
    T. Huynh-The is with the ICT Convergence Research Center, Kumoh National Institute of Technology, 
    Korea.
    W.-J. Hwang is with the Department of Biomedical Convergence Engineering, Pusan National University, 
    Korea.
    E-mails: \{vietpq@pusan.ac.kr, mzeng@mun.ca, thienht@kumoh.ac.kr, ruby@szu.edu.cn, wjhwang@pusan.ac.kr\}.}
    \thanks{This work was supported by a National Research Foundation of Korea (NRF) Grant funded by the Korean Government (MSIT) under Grants NRF-2019R1C1C1006143 and NRF-2019R1I1A3A01060518.}
}

\maketitle

\begin{abstract}
Federated learning (FL), invented by Google in 2016, has become a hot research trend. However, enabling FL in wireless networks has to overcome the limited battery challenge of mobile users. 
In this regard, we propose to apply unmanned aerial vehicle (UAV)-empowered wireless power transfer to enable sustainable FL-based wireless networks. The objective is to maximize the UAV transmit power efficiency, via a joint optimization of transmission time and bandwidth allocation, power control, and the UAV placement. Directly solving the formulated problem is challenging, due to the coupling of variables. 
Hence, we leverage the decomposition technique and a successive convex approximation approach to develop an efficient algorithm, namely UAV for sustainable FL (UAV-SFL).
Finally, simulations illustrate the potential of our proposed UAV-SFL approach in providing a sustainable solution for FL-based wireless networks, and in reducing the UAV transmit power by 32.95\%, 63.18\%, and 78.81\% compared with the benchmarks.   
\end{abstract}


\IEEEpeerreviewmaketitle

\section{Introduction}
\label{Sec:Introduction}
\IEEEPARstart{T}{o} overcome challenges of data privacy in conventional deep learning, the concept of federated learning (FL) has been invented by Google in 2016 \cite{konevcny2016federatedlearning}. In FL, users need not share their data with the central server but only information of the local models is transmitted to the server. 
Thanks to its characteristics, FL has been adopted in many applications such as Google keyboard suggestions \cite{yang2018applied}, medicine \cite{sheller2020federated}, and mobile communications \cite{Liu2020FederatedL6G}. 
Although FL has found many benefits in enabling privacy-preserving learning solutions, it also poses significant challenges such as resource management, robustness, and incentive mechanisms \cite{lim2020federated}. 
Moreover, powering mobile users with limited battery capacity is of vital importance to enable FL as users may run out of battery during the training process. In this regard, wireless powered communication networks (WPCNs) represent a promising solution, owing to their ability to charge the users wirelessly. 
Nonetheless, conventional WPCNs often suffer from several challenges, such as doubly near-far issue (i.e., far users need to spend higher power with smaller harvested energy) and severe performance degradation over long distance  \cite{xie2019throughput}. To address these challenges, unmanned aerial vehicle (UAV) enabled WPCNs have been introduced, which often supports line-of-sight connections with flexible deployment and controllable mobility \cite{xie2019throughput}, thus improving the network performance.

Over the last two years, various studies have been dedicated to 1) FL-WPCNs \cite{tran2020lightwave} and 2) FL-UAV communications \cite{liu2020federated, wang2020learning, shiri2020communication}. As the very first work on FL-WPCNs, Tran \etal \cite{tran2020lightwave} considered the application of lightwave (i.e., visible and infrared light) power transfer to wirelessly power mobile users, which then utilize the harvested power to upload their local models to the learning server over radio channels. This work illustrated that users are replenished with enough energy to perform FL computation. However, the framework in \cite{tran2020lightwave} requires synchronizations among users, which is a  challenging assumption as users may have different computing capabilities and data sizes. With regard to FL-UAV communications, the works in \cite{liu2020federated, wang2020learning, shiri2020communication} employed FL to enable UAV applications such as quality sensing with UAV swarms \cite{liu2020federated}, secure UAV crowdsensing with blockchain \cite{wang2020learning}, and massive UAV communications \cite{shiri2020communication}. 

Despite promising results, we are not aware of any work that focuses UAV-enabled wireless-powered FL networks (UAV-WPFNs).
Such UAV-WPFNs are very promising in circumstances (e.g., large farms, mountains, and hard-to-reach areas), where a number of users 
with a limited battery capacity are deployed while no central server is available for data collection and model training.
In this work,
we propose to leverage the UAV to wirelessly power FL users so as to enhance the sustainability of FL-based networks. We aim to maximize the transmit power efficiency of the UAV by jointly optimizing transmission time allocation, bandwidth allocation, power control, and UAV placement. To address the formulated non-convex problem, we propose an iterative algorithm, namely UAV-SFL to achieve the maximum power efficiency of the UAV. At each iteration, the time allocation and the UAV transmit power are updated using the analogy of the bi-section rule, whereas power and bandwidth allocation of the users, and the UAV placement are updated by solving a convex feasibility problem. Simulation results show the sustainability of the considered UAV wireless-powered FL network, and its superiority over the benchmarks. 

\section{Problem Formulation}
\label{Sec:SystemModel}
\subsection{Network Model}
We consider a network setup with a rotary-wing single-antenna UAV and $K$ single-antenna users (e.g., mobile users, sensors, and Internet-of-Things (IoT) devices), which are randomly distributed within an area of interest. We denote the set of users as $\mathcal{K}=\{1,\dots,K\}$.
The UAV equipped with a mobile edge computing (MEC) server provides computing services and broadcasts energy to the users. Each user has an on-board computing processor to train the local model using its own data and has an energy circuit to harvest energy transmitted by the UAV. 
Similar to \cite{zhou2018computation}, we assume that the UAV can simultaneously broadcast energy and receives the local models from users. Meanwhile, each user can simultaneously harvest energy from the UAV, perform FL tasks, and communicate with the server.   

The rotary-wing UAV can hover at a given altitude $H$ while its coordinate can be optimized to establish LoS connections and maximize the network performance. Denote by $\mathbf{q} = \{x,y\}$ and $\mathbf{q}_{k} = \{x_{k},y_{k}\}$ the respective coordinates of the UAV and user $k$. The channel gain $g_{k}$ of user $k$ can be calculated as $g_{k} = \hat{\beta}_{0}(d_{k}/d_{0})^{-\alpha}$, 
where $\hat{\beta}_{0}$ is the reference channel gain at $d_{0} = 1$ m
and $d_{k}$ is the distance to the UAV. Here, the distance $d_{k}$ is given as $d_{k} = \sqrt{H^2 + \Vert \mathbf{q}-\mathbf{q}_{k} \Vert_{2}^{2}}$. 

The energy of user $k$ harvested from the UAV can be given as $E_{k} = \eta_{0}TPg_{k}$,
where $T$ is the time duration, $P$ is the transmit power of the UAV, and $0 < \eta_{0} \leq 1$ is the energy conversion efficiency of user $k$. This linear model has been adopted in many works (e.g. \cite{zhou2018computation} and \cite{yang2019energy}) to show meaningful insights of UAV-enabled wireless powered communications. The non-linear energy harvesting model, as considered in our work \cite{nguyen2021resource}, is an interesting topic for future work.

\subsection{Local Computing Model}
Each user has a local dataset, which is used to train the local model and is not shared with the server to protect user privacy. About local computation of user $k$, we denote as $f_{k}$ (in CPU cycles per second) the CPU computing capability, $D_{k}$ as the number of data samples, and $C_{k}$ as the number of CPU cycles needed to process a data sample. The computation time for one local iteration is $t_{k}^{\cp} = C_{k}D_{k} / f_{k}$.
The corresponding energy consumption is $E_{k}^{\cp} = \zeta_{k}C_{k}D_{k}f_{k}^{2}$,
where $\zeta_{k}$ is a coefficient depending on the hardware and chip architecture \cite{pham2019coalitional}.

\subsection{Communication Model}
After local computation, each user uploads its local model to the server for aggregation. We adopt frequency division multiple access (FDMA) for the uplink operation because FL with FDMA can be implemented in asynchronous manner \cite{yang2020energyFL, chen2020ajoint}. Other multiple access techniques such as rate splitting multiple access (RSMA) and non-orthogonal multiple access (NOMA) can be also applied with the consideration of synchronizations among users. 

For user $k$, the achievable rate is given as $R_{k} = b_{k}\log_{2}\left(1 + {p_{k}g_{k}}/{b_{k}n_{0}}\right)$,
where $b_{k}$ is the allocated bandwidth, $p_{k}$ is the transmit power, and $n_{0}$ is the noise power spectral density. 
Denote by $s$ the constant data size of the local model parameter and gradient that user $k$ needs to upload to the server. To ensure the transmission of $s$ within the uploading time $t_{k}^{\cm}$, the constraint $s \leq R_{k}t_{k}^{\cm}$ should hold. The corresponding energy consumed by user $k$ is $E_{k}^{\cm} = t_{k}^{\cm}p_{k}$. 

\subsection{Problem Formulation}
We aim at developing a resource allocation framework to \textcolor{black}{maximize} the power efficiency of the UAV so as to enable sustainable FL systems. In particular, we are interested in the following optimization problem.
\begin{subequations}
\label{OptPrb}
	\begin{align} 
	& \underset{P,\mathbf{p},\mathbf{f},\mathbf{b},\mathbf{t},\mathbf{q}}{\min} 
	& & P \label{OptPrb:obj}\\
	& \text{s.t.} 
	& & t_{k}^{\cm}b_{k}\log_{2}\left(1 + \frac{p_{k}g_{k}}{b_{k}n_{0}}\right) \geq s, \forall k \in \mathcal{K}, \label{OptPrb:rate}\\
	&&& N_{k}E_{k}^{\cp} + E_{k}^{\cm} \leq E_{k}, \forall k \in \mathcal{K}, \label{OptPrb:energy}\\
	&&& 0 \leq P \leq P^{\max}, \label{OptPrb:powerUAV}\\
	&&& 0 \leq p_{k} \leq p_{k}^{\max}, \forall k \in \mathcal{K}, \label{OptPrb:power}\\
	&&& \sum\nolimits_{k \in \mathcal{K}}b_{k} \leq B, \label{OptPrb:bAllocation}\\
	&&& t_{k}^{\cp}N_{k} + t_{k}^{\cm} \leq T, \forall k \in \mathcal{K}, \label{OptPrb:time}\\
	&&& f_{k}^{\min} \leq f_{k} \leq f_{k}^{\max}, \forall k \in \mathcal{K}, \label{OptPrb:CPUfre}\\
	&&& \lVert \mathbf{q} \rVert_{2}^{2} \triangleq x^2 + y^2 \leq C^2, \label{OptPrb:coordinate}
	\end{align}
\end{subequations}
where $\mathbf{p} = \{p_{1},\dots,p_{K}\}$, $\mathbf{f} = \{f_{1},\dots,f_{K}\}$, $\mathbf{b} = \{b_{1},\dots,b_{K}\}$, and $\mathbf{t} = \{t_{1}^{\cm},\dots,t_{K}^{\cm}\}$. 
In the above problem, \eqref{OptPrb:rate} indicates the constraint on the data rate transmission, \eqref{OptPrb:energy} implies that the energy harvested from the UAV should be greater than the energy consumed for local computation and communication, \eqref{OptPrb:powerUAV} and \eqref{OptPrb:power} indicate the feasible range of the transmit power due to power budgets of the UAV and users, and~\eqref{OptPrb:bAllocation} denotes the bandwidth constraint with $B$ being the system bandwidth. Constraint~\eqref{OptPrb:time} ensures that each global round should be completed within the time frame $T$, with $N_{k}$ being the number of local iterations required by each user. We do emphasize that $N_{k}$ can be preset, and the value is decided by the target accuracy of the local model \cite{yang2020energyFL,tran2020lightwave}. The CPU frequency of each user is constrained in~\eqref{OptPrb:CPUfre} with $f_{k}^{\min}$ and $f_{k}^{\max}$ being the minimum and maximum CPU frequency of user $k$, respectively. Finally, \eqref{OptPrb:coordinate} expresses that the UAV should be placed within a circular disc with radius $C$. 
Solving problem~\eqref{OptPrb} directly is challenging since multiple optimization variables are coupled. Moreover, the optimization of UAV location arises the non-linearity of~\eqref{OptPrb} caused by the channel gain model and the $\log_{2}(\cdot)$ function of the achievable rates. 

\section{Proposed Solution}
\label{Sec:SolutionApproaches}

To overcome the non-linearity due to the UAV placement we introduce auxiliary variables $\mathbf{u}_{k} = \{u_{k}\}_{k \in \mathcal{K}}$ and rewrite the problem~\eqref{OptPrb} as follows:
\begin{subequations}
\label{OptPrb1}
	\begin{align} 
	& \underset{P,\mathbf{p},\mathbf{f},\mathbf{b},\mathbf{t},\mathbf{q},\mathbf{u}}{\min} 
	& & P \label{OptPrb1:obj}\\
	& \text{s.t.} 
	& & t_{k}b_{k}\log_{2}\left(1 + \frac{p_{k}g_{0}}{b_{k}u_{k}}\right) \geq s, \forall k \in \mathcal{K}, \label{OptPrb1:rate}\\
	&&& N_{k}\zeta_{k}C_{k}D_{k}f_{k}^{2} + t_{k}p_{k} \leq \frac{P\beta_{0}}{u_{k}}, \forall k \in \mathcal{K}, \label{OptPrb1:energy}\\
	&&& \frac{N_{k}C_{k}D_{k}}{f_{k}} + t_{k} \leq T, \forall k \in \mathcal{K}, \label{OptPrb1:time}\\
	&&& u_{k} \geq \left(H^2 + \Vert \mathbf{q}-\mathbf{q}_{k} \Vert_{2}^{2}\right)^{\alpha/2}, \forall k \in \mathcal{K}, \label{OptPrb1:aux_vars}\\
	&&& 
	\eqref{OptPrb:powerUAV}, \eqref{OptPrb:power}, \eqref{OptPrb:bAllocation}, 
	\eqref{OptPrb:CPUfre}, \eqref{OptPrb:coordinate},
	\end{align}
\end{subequations}
where $\beta_{0} = \eta_{0}T\hat{\beta}_{0}$ and $g_{0} = \hat{\beta}_{0}/n_{0}$.
We also replace $t_{k}^{\cm}$ by $t_{k}$ for ease of presentation. It is however still difficult to solve the problem~\eqref{OptPrb1}. In this work, we leverage the block successive upper-bound minimization (BSUM) method proposed in \cite{hong2016unified} to solve~\eqref{OptPrb1}. Particularly, we divide the entire variables of the problem~\eqref{OptPrb1} into three blocks: $(\mathbf{t},\mathbf{f})$, $P$, and $(\mathbf{p},\mathbf{b},\mathbf{q},\mathbf{u})$, and as will be shown in Algorithm~\ref{Alg:BSUM_SFL}, these blocks can be updated alternately in an iterative manner.

For the first block variable $(\mathbf{t},\mathbf{f})$, optimizing the problem~\eqref{OptPrb1} is equivalent to minimizing the energy consumed for both local computation and communication. 
%
From~\eqref{OptPrb1:rate},
the minimum transmission time of user $k$ at the update step $\kappa$ is given as
\begin{equation} \label{Eq:minTime}
    t_{k}^{\min,(\kappa)} = \frac{s}{b_{k}\log_{2}\left(1 + p_{k}g_{0} / b_{k}u_{k}\right)}.
\end{equation}
Also from~\eqref{OptPrb1:time},
at the first update step (i.e. $\kappa = 0$), the transmission time of user $k$ is limited by the following upper-bound value
\begin{equation} \label{Eq:maxTime}
    t_{k}^{\max,(\kappa)} = T - \frac{N_{k}C_{k}D_{k}}{f_{k}^{\max}}.
\end{equation}
Therefore, the transmission time of user $k$ is a feasible point within the range $[t_{k}^{\min,(\kappa)}, t_{k}^{\max,(\kappa)}]$. 
%
As the transmission energy of each user becomes smaller when the transmission time is smaller 
(as $E_{k}^{\cm} = t_{k}p_{k}$)
and there is one additional constraint~\eqref{OptPrb1:energy},
the transmission time can be updated using the analogy of the bi-section rule. In particular, at the update step $\kappa$, the transmission time of user $k$ is set as follows:
\begin{equation} \label{Eq:Update_t}
    t_{k}^{(\kappa)} = ({t_{k}^{\max,(\kappa)} + t_{k}^{\min,(\kappa)}})/{2}\textcolor{blue}{,}
\end{equation}
and the maximum transmission time of user $k$ for the next update step is set as $t_{k}^{\max,(\kappa+1)} = t_{k}^{(\kappa)}$, 
Here, we should note that constraint~\eqref{OptPrb1:energy}
of the block variable $(\mathbf{t},\mathbf{f})$ is handled via the optimization of the other blocks.
For a given transmission time $t_{k}^{(\kappa)}$, we can infer from \eqref{OptPrb:CPUfre} and \eqref{OptPrb1:time}
that the optimal CPU frequency of user $k$ should satisfy $f_{k}^{(\kappa)} = \min\{\max\{\underline{f}_{k}^{\min},f_{k}^{\min}\},f_{k}^{\max}\}$ in order to minimize the energy consumed for local computation, where 
\begin{equation} \label{Eq:Update_fmin}
    \underline{f}_{k}^{\min} = \frac{N_{k}C_{k}D_{k}}{T - t_{k}^{(\kappa)}}.
\end{equation}

Similarly for the block of the transmit power $P$ of the UAV, at the first update step, the maximum transmit power is set as $P^{\max}_{(\kappa = 0)} = P^{\max}$ and the minimum power can be derived from~\eqref{OptPrb1:energy}
as $P^{\min}_{(\kappa)} = \max\{P_{k}^{(\kappa)}\}$, where
\begin{equation}
P_{k}^{(\kappa)}\ = \frac{u_{k}}{\beta_{0}}\left({\color{black}N_{k}}\zeta_{k}C_{k}D_{k}f_{k}^{2} + t_{k}p_{k}\right). 
\end{equation}
Similar to the update rule of the transmission time, the transmit power of the UAV can be updated as follows:
\begin{equation} \label{Eq:Update_P}
    P^{(\kappa)} = ({P^{\max}_{(\kappa)} + P^{\min}_{(\kappa)}})/{2}.
\end{equation}
From~\eqref{Eq:Update_P}, the transmit power of the UAV decreases over the course of optimization. The reason is that the energy consumed for both local computation and transmission gets smaller, as optimized in the first block phase.
And the maximum transmit power of the UAV for the next update step is set as $P^{\max}_{(\kappa+1)} = P^{\min}_{(\kappa)}$.

For given $(\mathbf{t},\mathbf{f},P)$, the problem can be equivalently represented as the following feasibility problem:
\begin{subequations}
\label{OptPrb2}
	\begin{align} 
	& \find  
	& & \{\mathbf{p},\mathbf{b},\mathbf{q},\mathbf{u}\} \label{OptPrb2:obj}\\
	& \text{s.t.} 
	& & b_{k}\log_{2}\left(1 + \frac{p_{k}g_{0}}{b_{k}u_{k}}\right) \geq s/t_{k}, \forall k \in \mathcal{K}, \label{OptPrb2:rate}\\
	&&& {\color{black}N_{k}}\zeta_{k}C_{k}D_{k}f_{k}^{2} + t_{k}p_{k} \leq \frac{P\beta_{0}}{u_{k}}, \forall k \in \mathcal{K}, \label{OptPrb2:energy}\\
	&&& u_{k} \geq \left(H^2 + \Vert \mathbf{q}-\mathbf{q}_{k} \Vert_{2}^{2}\right)^{\alpha/2}, \forall k \in \mathcal{K}, \label{OptPrb2:auxvar}\\
	&&& \eqref{OptPrb:power}, \eqref{OptPrb:bAllocation}, \eqref{OptPrb:coordinate}.
	\end{align}
\end{subequations}
This problem is non-convex, but the feasible set composed of \eqref{OptPrb:power}, \eqref{OptPrb:bAllocation}, \eqref{OptPrb:coordinate}, and~\eqref{OptPrb2:auxvar} is convex. To convexify~\eqref{OptPrb2:rate}, we introduce the following lemma, and its proof can be found in \cite{nasir2019uav}. 
\begin{lemma}
For every $ x > 0 $, $ y > 0 $, $\tau > 0$, $ \bar{x} > 0 $, $ \bar{y} > 0 $, $ \bar{\tau} > 0 $, we have the following inequality
\begin{align}
\tau\ln\left(1 + \frac{1}{xy}\right) \geq 
& 2\bar{\tau}\ln\left(1 + \frac{1}{\bar{x}\bar{y}}\right) + \frac{\bar{\tau}}{1 + \bar{x}\bar{y}}\left(2 - \frac{x}{\bar{x}} - \frac{y}{\bar{y}}\right) \notag\\
& - \frac{\bar{\tau}^{2}\ln(1 + 1/\bar{x}\bar{y})}{\tau}. \label{Ineq}
\end{align}
\end{lemma}

With regard to constraint~\eqref{OptPrb2:rate}, at the update step $\kappa$, applying the inequality~\eqref{Ineq} with $\tau = b_{k}, x = u_{k}/p_{k}g_{0}, y = b_{k}, \bar{\tau} = b_{k}^{(\kappa-1)}, \bar{x} = u_{k}^{(\kappa-1)}/p_{k}^{(\kappa-1)}g_{0}, \bar{y} = b_{k}^{(\kappa-1)}$ yields
\begin{align} \label{Ineq:Rate}
& b_{k}\log_{2}\left(1 + \frac{p_{k}g_{0}}{b_{k}u_{k}}\right) \geq \notag\\ & \; \frac{1}{\ln(2)} \left( \lambda_{k} + \mu_{k}\left( 2 - \frac{u_{k}}{p_{k}}\frac{p_{k}^{(\kappa-1)}}{u_{k}^{(\kappa-1)}} - \frac{b_{k}}{b_{k}^{(\kappa-1)}} \right) - \frac{\nu_{k}}{b_{k}} \right),
\end{align}
with 
\begin{align}
& \lambda_{k} = 2b_{k}^{(\kappa-1)}\ln\left(1 + \frac{p_{k}^{(\kappa-1)}g_{0}}{b_{k}^{(\kappa-1)}u_{k}^{(\kappa-1)}}\right), \\
& \mu_{k} = b_{k}^{(\kappa-1)}\left(1 + \frac{b_{k}^{(\kappa-1)}u_{k}^{(\kappa-1)}}{p_{k}^{(\kappa-1)}g_{0}}\right)^{-1}, \\
& \nu_{k} = (b_{k}^{(\kappa-1)})^{2}\ln\left(1 + \frac{p_{k}^{(\kappa-1)}g_{0}}{b_{k}^{(\kappa-1)}u_{k}^{(\kappa-1)}}\right).
\end{align}
Further, we have the following inequality
\begin{align}
& \frac{u_{k}}{p_{k}}\frac{p_{k}^{(\kappa-1)}}{u_{k}^{(\kappa-1)}}  \notag\\
& = \frac{1}{4} \left(\left(\frac{u_{k}}{u_{k}^{(\kappa-1)}} + \frac{p_{k}^{(\kappa-1)}}{p_{k}}\right)^{2} - \left(\frac{u_{k}}{u_{k}^{(\kappa-1)}} - \frac{p_{k}^{(\kappa-1)}}{p_{k}}\right)^{2}\right) \notag\\
& \leq \frac{1}{4} \left(\frac{u_{k}}{u_{k}^{(\kappa-1)}} + \frac{p_{k}^{(\kappa-1)}}{p_{k}}\right)^{2} \triangleq \pi_{k}^{(\kappa)}(p_{k},u_{k}).
\end{align}
Therefore, \eqref{Ineq:Rate} can be transformed into the following
\begin{align} \label{Ineq:quad1}
& b_{k}\log_{2}\left(1 + \frac{p_{k}g_{0}}{b_{k}u_{k}}\right) \notag\\
& \geq \frac{1}{\ln(2)} \left( \lambda_{k} + \mu_{k}\left( 2 - \pi_{k}^{(\kappa)}(p_{k},u_{k}) - \frac{b_{k}}{b_{k}^{(\kappa-1)}} \right) - \frac{\nu_{k}}{b_{k}} \right) \notag\\
& \triangleq R_{k}^{(\kappa)}(b_{k},p_{k},u_{k}).
\end{align}

With regard to~\eqref{OptPrb2:energy}, since both sides are convex expressions, this constraint is non-convex. 
Fortunately, we can leverage the first-order optimality condition to approximate the right hand side of~\eqref{OptPrb2:energy} as follows \cite{boyd2004convex}: 
\begin{equation}
\frac{P\beta_{0}}{u_{k}} \geq  \frac{P\beta_{0}}{u_{k}^{(\kappa-1)}} - \frac{P\beta_{0}}{\left(u_{k}^{(\kappa-1)}\right)^2}\left(u_{k} - u_{k}^{(\kappa-1)}\right) \triangleq \phi_{k}^{(\kappa)}(u_{k}). 
\end{equation} 
As a result, at the iteration $\kappa$, constraint~\eqref{OptPrb2:energy} of user $k$ can be approximated as 
\begin{equation} \label{InEq:energy_appro}
{{\color{black}N_{k}}\zeta_{k}C_{k}D_{k}f_{k}^{2}} + t_{k}p_{k} \leq \phi_{k}^{(\kappa)}(u_{k}).
\end{equation}
To summarize, we solve the following convex problem to update the block variable $(\mathbf{p}^{(\kappa)},\mathbf{b}^{(\kappa)},\mathbf{q}^{(\kappa)},\mathbf{u}^{(\kappa)})$.
\begin{subequations}
\label{OptPrb3}
	\begin{align} 
	& \find  
	& & \{\mathbf{p},\mathbf{b},\mathbf{q},\mathbf{u}\} \label{OptPrb3:obj}\\
	& \text{s.t.} 
	& & R_{k}^{(\kappa)}(b_{k},p_{k},u_{k}) \geq s/t_{k}, \forall k \in \mathcal{K}, \label{OptPrb3:rate}\\
	&&& {{\color{black}N_{k}}\zeta_{k}C_{k}D_{k}f_{k}^{2}} + t_{k}p_{k} \leq \phi_{k}^{(\kappa)}(u_{k}), \forall k \in \mathcal{K}, \label{OptPrb3:energy}\\
	&&& \eqref{OptPrb:power}, \eqref{OptPrb:bAllocation}, \eqref{OptPrb:coordinate}, 
	\eqref{OptPrb2:auxvar}.
	\end{align}
\end{subequations}

\begin{algorithm}[t]
	\caption{Proposed Algorithm}\label{Alg:BSUM_SFL}
	\begin{algorithmic}[1]
		\State Set the iteration index $\kappa = 0$ and {initialize} a feasible solution $(P^{(\kappa)}, \mathbf{p}^{(\kappa)}, \mathbf{f}^{(\kappa)}, \mathbf{b}^{(\kappa)}, \mathbf{t}^{(\kappa)}, \mathbf{q}^{(\kappa)})$ for the problem~\eqref{OptPrb}.
		
		\Repeat
		\State Set $\kappa \leftarrow \kappa + 1$.
		
		\State Update $t_{k}^{(\kappa)}$ according to~\eqref{Eq:Update_t}.
		
		\State Update $f_{k}^{(\kappa)}$ with $\underline{f}_{k}^{\min}$ in~\eqref{Eq:Update_fmin}.
		
		\State Update $P$ based on~\eqref{Eq:Update_P}.
		
		\State Solve~\eqref{OptPrb3} to update $(\mathbf{p}^{(\kappa)}, \mathbf{b}^{(\kappa)}, \mathbf{q}^{(\kappa)})$.
		
		\Until{$|P^(\kappa) - P^(\kappa-1)|/P^(\kappa-1) \leq \epsilon$.}
	\end{algorithmic}
\end{algorithm}

The proposed algorithm is summarized in Algorithm~\ref{Alg:BSUM_SFL}. In each iteration, block variables are updated cyclically. The algorithm stops when the relative tolerance is less than a preset threshold $\epsilon$. As the transmit power of the UAV is reduced after each update step, Algorithm~\ref{Alg:BSUM_SFL} converges to the final solution after a number of steps. At each step, the computational complexity of solving the convex problem~\eqref{OptPrb3} is $\mathcal{O}((3K + 2)^{2}(4K + 2)^{2.5} + (4K + 2)^{3.5})$ as it involves $4K + 2$ constraints and $3K + 2$ variables \cite{nasir2019uav}.

\section{Simulation Results}
\label{Sec:Simulations}
For validating our proposed method, namely UAV-SFL, we deploy a network, in which $K=25$ users are randomly distributed within a circular disc with a radius of $50$ m, and the rotary-wing UAV hovers at a preset altitude of $20$ m. The system bandwidth $B = 20$ MHz, the respective maximum powers of users and the UAV are $10$ dBm and $36$ dBm,
and the energy harvesting efficiency $\eta_{0} = 0.9$. For local computation, the maximum and minimum CPU frequencies $f_{k}^{\max} = 1$ GHz and $f_{k}^{\min} = 0.1$ GHz for all users, the data size $s = 100$ Kb, the coefficient of chip $\zeta = 10^{-28}$, the local data $D_{k}$ and the workload $C_{k}$ are randomly selected from the ranges $[5,10]$ Mb and $[10, 20]$ cycles per bit, the maximum number of local iterations $N_{k} = 4$, and the time frame $T = 8$ s. 

\begin{figure}[t]
	\centering
	\includegraphics[width=0.950\linewidth]{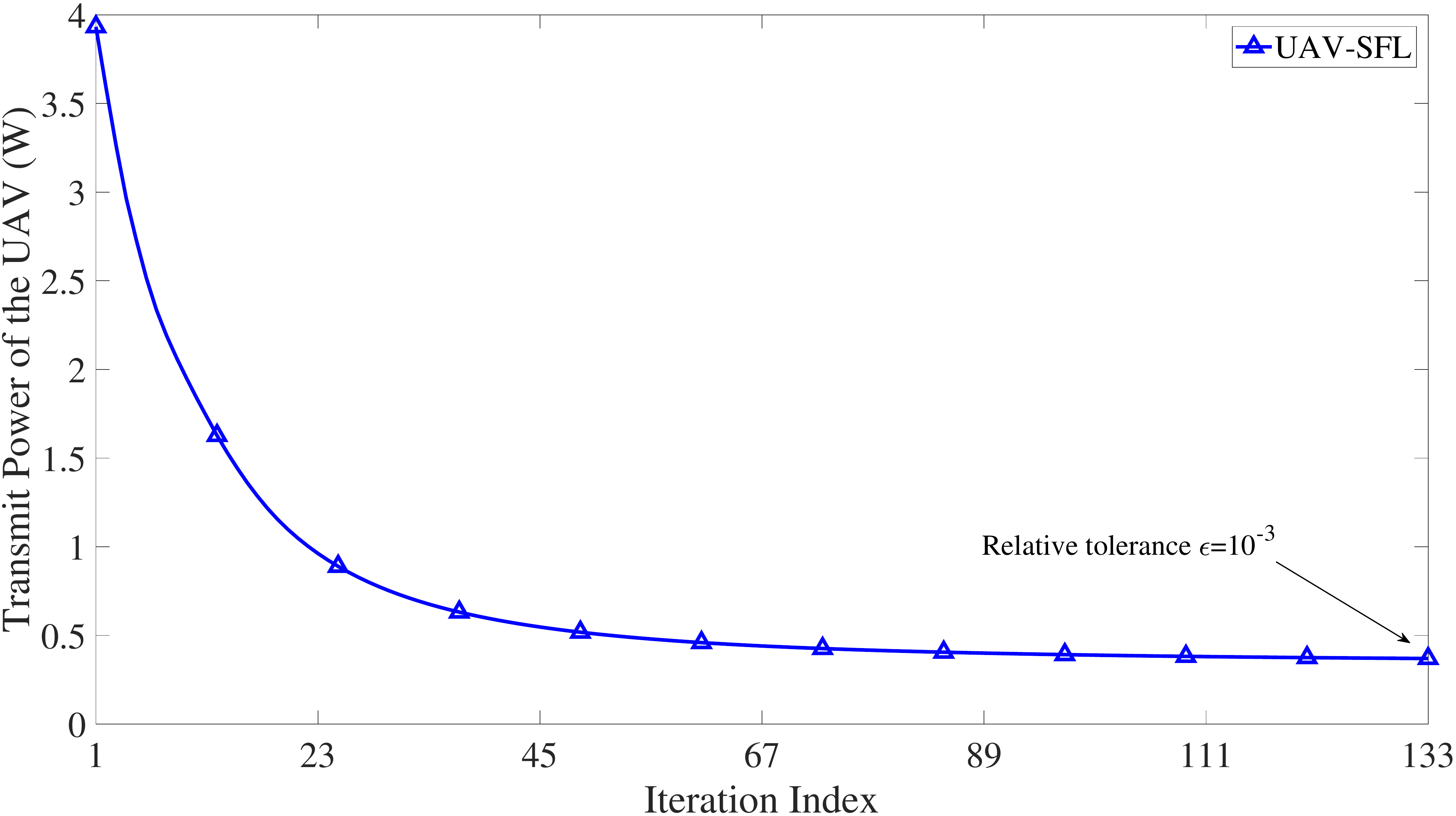}
	\caption{Convergence evolution of the proposed algorithm.}
	\label{Fig:Convergence}
\end{figure}
Firstly, we present the convergence evolution of UAV-SFL in Fig.~\ref{Fig:Convergence}. This figure shows that Algorithm~\ref{Alg:BSUM_SFL} can converge to the final solution after a reasonable number of iterations. We recall that at each step, we solve the convex feasibility problem~\eqref{OptPrb3}, and further update time allocation, transmit power of the UAV, and CPU frequencies using linear computations. Consequently, the total complexity of our proposed UAV-SFL algorithm is very reasonable. 


\begin{figure}[t]
	\centering
	\subfloat[\label{Fig:Time_vs_Nk}]{\includegraphics[width=0.470\linewidth]{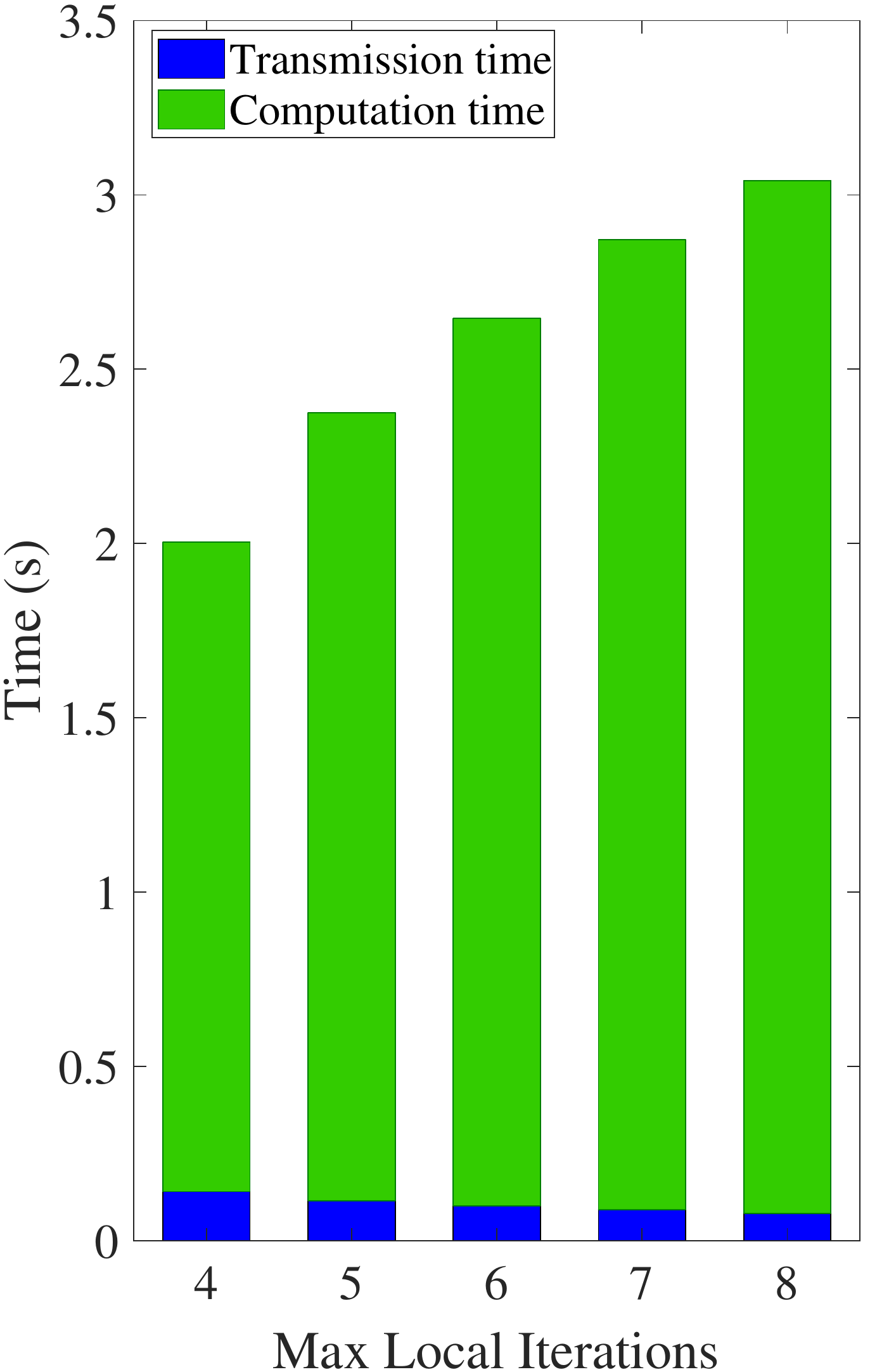}} \;\;\;
	\subfloat[\label{Fig:UAVPower_vs_Nk}]{\includegraphics[width=0.475\linewidth]{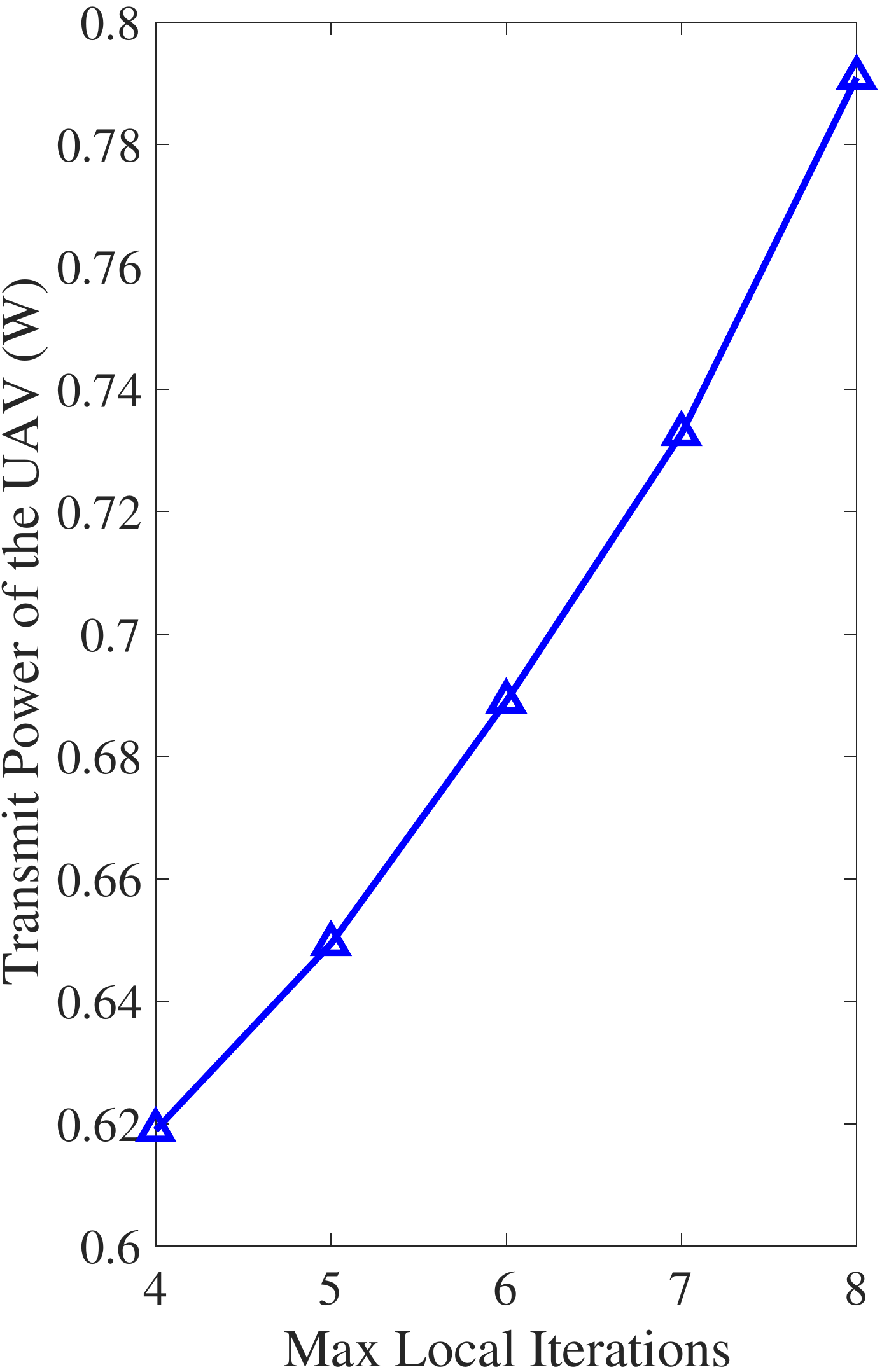}}
	\caption{Performance vs. maximum number of local iterations. (a) computation and transmission time. (b) transmit power of the UAV.}
	\label{Fig:Performance_vs_Nk}
\end{figure} 

Secondly, in Fig.~\ref{Fig:Time_vs_Nk}, we show the average transmission and computation time when varying the maximum number of local iterations. The figure indicates that the higher the maximum number of local iteration is, the larger the computation time is, while the smaller the transmission time is. This is reasonable since the computation time is directly proportional to $N_k$, while the total completion time is required to be less than the time frame $T$, as indicated in \eqref{OptPrb:time}. We can also observe from Fig.~\ref{Fig:UAVPower_vs_Nk} that the transmit power of the UAV increases as $N_{k}$ increases. 
The reason is that the majority of the total harvested energy is for executing local computation, and thus the UAV should increase its transmit power to ensure that each user can harvest enough energy to perform FL tasks. Therefore, the proposed UAV-SFL approach can adaptively adjust the transmit power of the UAV to wirelessly power users to complete their FL tasks within the time frame.

\begin{figure}[t]
	\centering
	\subfloat[\label{Fig:UAVPower_vs_size}]{\includegraphics[width=0.460\linewidth]{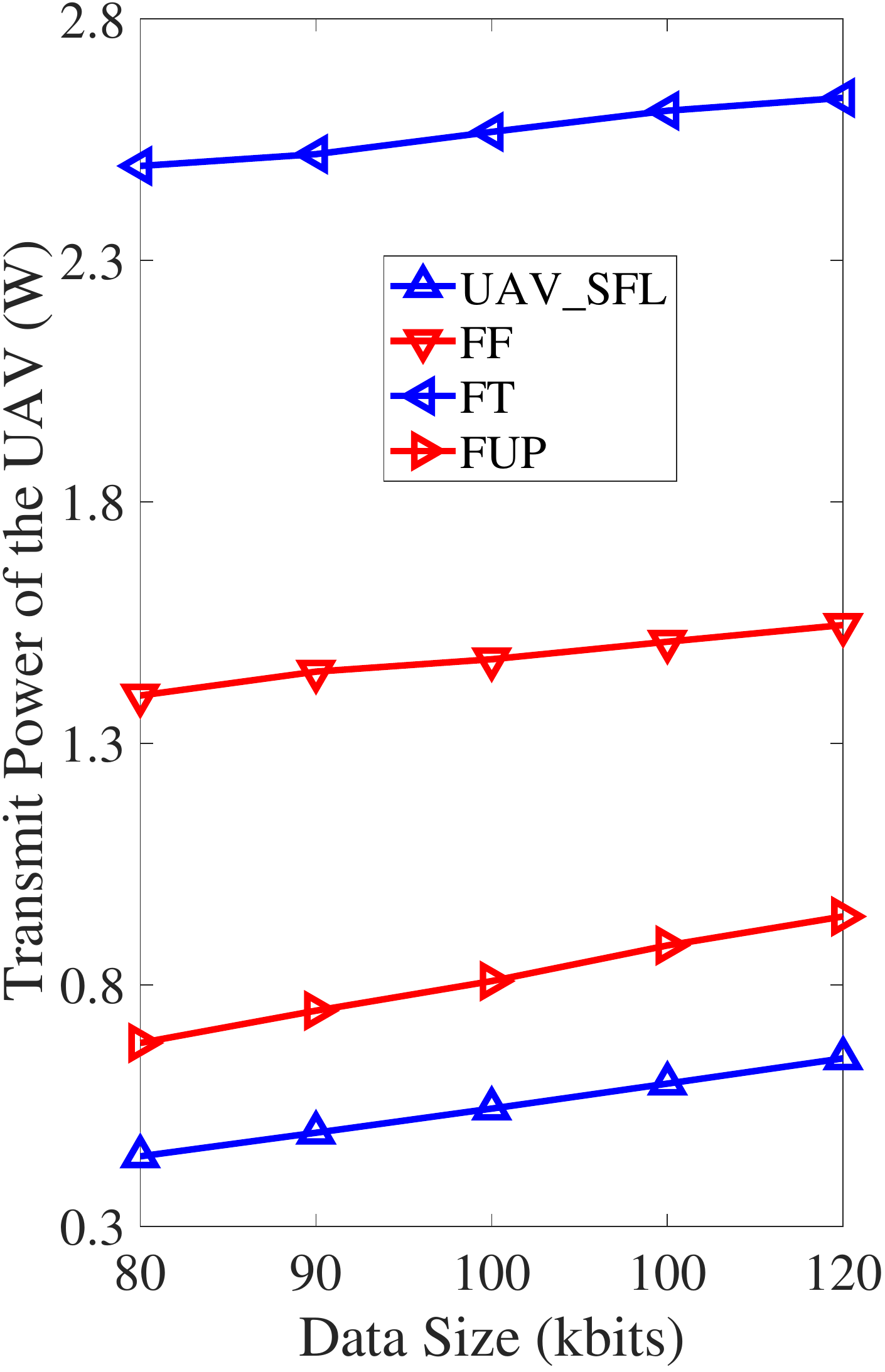}} \;\;\;
	\subfloat[\label{Fig:UAVPower_vs_B}]{\includegraphics[width=0.460\linewidth]{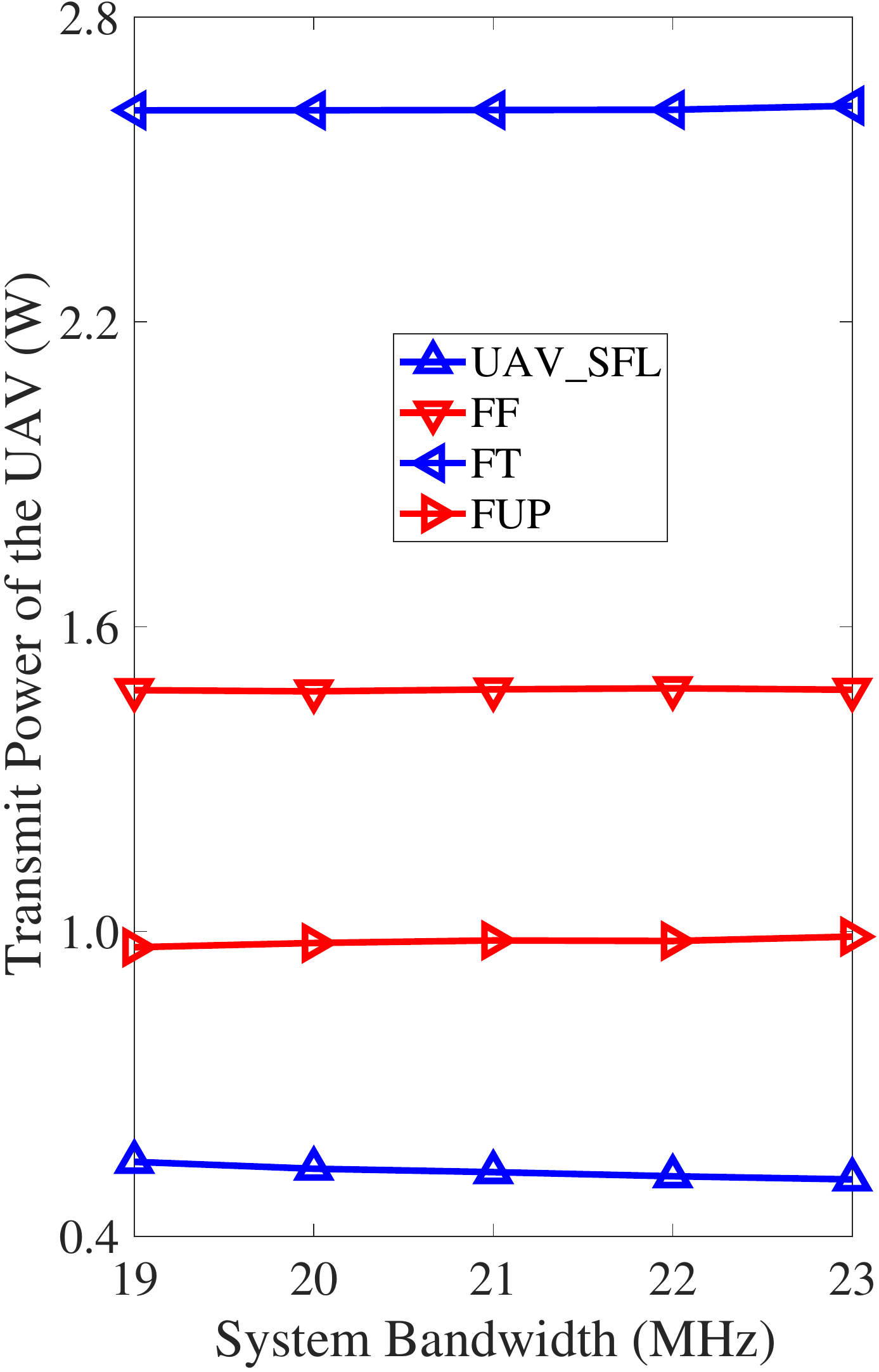}} 
	\caption{Comparison of UAV-SFL with the other approaches.}
	\label{Fig:Performance_Comparison}
\end{figure} 

Finally, we draw a comparison with three other approaches, including 1) a scheme with \underline{F}ixed CPU \underline{F}requencies (labeled as FF), 2) a scheme with \underline{F}ixed transmission \underline{T}ime (labeled as FT), and 3) a scheme with \underline{F}ixed \underline{U}AV \underline{P}lacement (labeled as FUP). From Fig.~\ref{Fig:UAVPower_vs_size}, the transmit power of the UAV increases as the data size $s$ increases. This is because the transmission time and users' transmit power should increase to complete the local model transmission, which consequently demands a higher transmit power of the UAV. The figure also shows that the transmit power of the UAV derived from our UAV-SFL approach via jointly optimizing the UAV placement, time, and CPU frequencies is significantly lower than those of three benchmarks. On average, UAV-SFL is better than FUP, FF, and FT by 32.95\%, 63.18\%, and 78.81\%, respectively. 
Fig.~\ref{Fig:UAVPower_vs_B} plots the transmit power of the UAV as a function of the system bandwidth. From this figure, there is just a minor change in the transmit power of the UAV as the system bandwidth increases. This is because the UAV does not change its optimized coordinate much when the system bandwidth changes \cite{nasir2019uav}. With regard to our UAV-SFL scheme, 
the transmit power of the UAV slightly reduces as the system bandwidth increases. The reason for this is that increasing the system bandwidth results in larger achievable rates, which reduces the energy needed for transmission and consequently decreases the transmit power of the UAV. Again, this figure shows the superiority of the proposed UAV-SFL approach over the benchmarks.

\section{Conclusion}
\label{Sec:Conclusion}
In this article, we have considered a network model, where the UAV is deployed to wirelessly power users so as to support sustainable FL-based wireless networks. We have developed an efficient algorithm to obtain the solution with a convergence guarantee. Presented simulation results show that the proposed scheme outperforms the benchmarks with fixed CPU frequencies, transmission time and UAV placement. Several promising directions can be stemmed from our current work such as consideration of other multiple-access techniques (e.g., NOMA and RSMA), fixed-wing UAVs, UAV-assisted visible light communications, and multi-UAV WPFNs. Moreover, as this work is considered for one global round, the extension to multiple global rounds is very promising.


\end{document}